# Title: Visualizing Broken Symmetry and Topological Defects in a Quantum Hall Ferromagnet

**Authors:** Xiaomeng Liu[1†], Gelareh Farahi[1†], Cheng-Li Chiu[1†], Zlatko Papic[2], Kenji Watanabe[3], Takashi Taniguchi[4], Michael P. Zaletel[5], Ali Yazdani[1*]

**Affiliations:**

[1] Joesph Henry Laboratories and Department of Physics, Princeton University, Princeton, NJ 08544, USA

[2] School of Physics and Astronomy, University of Leeds, Leeds LS2 9JT, UK

[3] Research Center for Functional Materials, National Institute for Materials Science, 1-1 Namiki, Tsukuba 305-0044, Japan

[4] International Center for Materials Nanoarchitectonics, National Institute for Materials Science, 1-1 Namiki, Tsukuba 305-0044, Japan

[5] Department of Physics, University of California at Berkeley, Berkeley, CA 94720, USA

*Correspondence to: yazdani@princeton.edu

† These authors contributed equally to this work

**Abstract:** The interaction between electrons in graphene under high magnetic fields drives the formation of a rich set of quantum Hall ferromagnetic phases (QHFM), with broken spin or valley symmetry. Visualizing atomic scale electronic wavefunctions with scanning tunneling spectroscopy (STS), we resolve microscopic signatures of valley ordering in QHFM and fractional quantum Hall phases of graphene. At charge neutrality, we observe a field-tuned continuous quantum phase transition from a valley polarized state to an intervalley coherent state, with a Kekule distortion of its electronic density. Mapping the valley texture extracted from STS measurements of the Kekule phase, we visualize valley skyrmion excitations localized near charged defects. Our techniques can be applied to examine valley ordered phases and their topological excitations in a wide range of materials.

**Main Text:**

Quantum Hall ferromagnets are broken symmetry states, in which the exchange interaction between electrons in Landau levels gives rise to quantum Hall phases with polarized or coherent superposition of spin, valley, or orbital degrees of freedom[1]. In the presence of a magnetic field, a variety of two-dimensional electronic systems, including those in semiconductors[1,2], graphene[2], and an increasing number of moiré flat bands materials host a diversity of QHFM phases[3–8]. Thus far, these interacting and topological phases of matter have been examined macroscopically, usually through study of their transport properties. However, the microscopic features of the electronic wavefunctions of these phases can directly reveal the nature of their broken symmetry[9,10] and more importantly determine the nature of the excitations they host. A particularly interesting aspect of broken symmetry states is their topological excitations, such as skyrmions[11–13], which not only determine the stability of such phases, but their interactions may also lead to the formation of more exotic quantum phases, such as those recently proposed in moiré materials[14–16].

Monolayer graphene's SU(4) isospin space consisting of spin and valley gives rise to a rich array of QHFM phases, which have been studied using transport and thermodynamic measurements[2]. Particularly intriguing is the electrically insulating phase at charge neutrality point (CNP) at high magnetic fields[17], since with two out of four isospin flavors occupied, spin and valley cannot be simultaneously polarized due to Pauli exclusion. Theoretical efforts have predicted a rich phase diagram of four possible broken symmetry QHFM states at charge neutrality[18], a charge density wave (CDW) phase which is sublattice and valley polarized and spin unpolarized, the spin ferromagnet (FM) which is a quantum spin Hall insulator, the canted anti-ferromagnet (CAF) in which spins on different sublattices point in the near-opposite directions, and intervalley coherent (IVC) state with a Kekule reconstruction, which is spin unpolarized. While transport studies have constrained aspect of the phase diagram[19,20], in the absence of microscopic measurements that probes the order parameter, nature of the ground state of graphene at charge neutrality has remained unresolved. Also unexplored are the plethora of topological excitations these phases have been predicted to host, such as a variety of skyrmions which may have complex flavor textures and even harbor fractional charge on the scale of the magnetic length[21–23]. Here we use spectroscopic mapping to visualize the broken symmetry states in graphene as a function of carrier concentration, including at charge neutrality, where we find evidence for localized valley skyrmions within the novel Kekule phase. Our work demonstrates the power of spectroscopic imaging to detect valley ordering and their topological excitations, which is applicable to a wider range of two-dimensional materials and their heterostructures.

The monolayer graphene devices used for our studies are fabricated on hexagonal Boron Nitride (hBN) substrates, with either graphite (devices A, C) or silicon back gates (device B) (see Fig. 1 for the experimental setup and an optical image of device A). Figure 1b&c show measurements of differential conductance dI/dV as function of sample bias $V_B$ measured over a wide range of filling factors $\nu$ ($\nu = 2\pi n l_B{}^2$, where $l_B = \sqrt{\hbar/eB}$ is the magnetic length and n the carrier density) controlled by the back gate voltage $V_g$. The Landau levels (LL) can be identified by their peaks in dI/dV, the energy spacing between which corresponds to the cyclotron energy $E_N = \hbar\omega_c\sqrt{N}$, where N is the LL orbital index, and $\hbar\omega_c\sim110$mV the cyclotron energy for B = 6T (Fig. 1d). As the filling factor increases, the Fermi energy is pinned within a LL as it is being filled and then jumps to the next LL at $\nu = \pm2, \pm6, \pm10$. For the incompressible states formed at these fillings, we find that energy gaps across the Fermi energy are enlarged by approximately a factor of two as compared to the expected cyclotron gap (Fig. S1). This effect is likely due to the graphene's bulk insulating behavior, when the chemical potential lies within these gaps (see discussion in the supplementary materials).

Electron-electron interaction driven symmetry breaking states are detected in our spectroscopic measurements as enlarged gaps at all the intermediate integer fillings, as shown in Fig. 1c. Despite numerous previous STS studies of graphene at high magnetic fields[24–34], our results are the first observation of symmetry breaking gaps in such experiments. The size of the gaps in our experiments are larger relative to those observed in transport and thermodynamic studies, a behavior similar to the incompressible insulating states at $\nu = \pm 2, \pm 6, \pm 10$. The absence of symmetry breaking gaps in previous STS measurements may be due the influence of tip-induced band bending or disorder, which we find to be negligible in our studies. First, our data as shown in Fig. 1c does not show any Coulomb diamond features associated with a tip-induced quantum dot, as seen in previous studies[29]. Second, we find that charge neutrality occurs near zero gate voltage, testifying that our sample is not doped by impurities and our measurements are not influenced by a tip-sample work function mismatch. Third, $V_B$ does not influence carrier density in the probed area since the dashed lines in Fig. 1c marking incompressible states are nearly vertical, therefore showing that tip gating is negligible. Finally, at partial fillings, the LLs are always pinned to the Fermi energy with their jumps aligned with the occurrence of the incompressible states, suggesting there is no density mismatch between the probed area and the bulk of the sample. It is possible that our tip effective radius is small compared to the magnetic length, so that the work-function mismatch between the tip and sample (which would typically lead to band-bending) traps at most zero or one electron charges below the tip, rather than  producing a well-defined change in filling factor in a larger region.

Beyond resolving the presence of broken symmetry states, our experiments also show direct signature of fractional quantum Hall (FQH) phases in spectroscopic measurements. Focusing on the STS properties between $v = -2$ and 2, as shown in Fig. 2a, we resolve enlarged gaps at partial filling of the zeroth LL (ZLL) corresponding to the fractional quantum Hall states at $v = \pm 4/3, \pm 2/3, \pm 1/3$. We corroborate the formation of FQH states in our devices by performing transport measurement while the STM tip height is reduced from the tunneling condition to directly contact the monolayer graphene (Fig. 2b). In this Corbino geometry, measurements of the conductance of our sample show dips at fractional fillings associated with the formation of FQH states. The observation of fractional states, up to 4/9[th] in our samples, at a modest magnetic field (6T) and at relatively elevated temperature (1.4K), attests to their high quality, making them comparable to the fully hBN encapsulated and dual graphite gated devices used for the highest quality transport measurements. Our ability to probe FQH phases in STM measurements paves the way to explore these topological phases and their exotic excitations in new ways, such as realization of methods for imaging anyons[35] or probing fractional edge states locally.

The spectroscopic measurements of the partially filled ZLL (Fig. 2a), including when the sample transitions through the FQH phases, always show splitting of the ZLL with a gap across the Fermi energy. This behavior is indicative of a Coulomb gap commonly observed when tunneling in and out of a two-dimensional electron gas at high magnetic fields[36–38]. The strong correlations among electrons in the flat LLs dictates that an additional energy is required for addition or removal of electrons from the system, resulting in a gap at the Fermi level that scales with the Coulomb energy $E_c = e^2/\epsilon l_B$, where $\epsilon$ is the effective dielectric constant. The field dependence of this gap at partial filling follows the expected $\sqrt{B}$ behavior, as shown in Fig. 2c, tracing Coulomb energy $E_c$ with a 0.62 scale factor, which corresponds well to those obtained from our exact diagonalization calculations (see supplementary materials).

To directly visualize broken valley symmetry of graphene's ZLL, we perform spectroscopic mapping of the electron and hole excitations of the ZLL (E-ZLL and H-ZLL respectively), with $V_B$ at the split ZLL peaks below or above the Coulomb gap. These spectroscopic dI/dV maps are performed with the STM tip at a constant height above the graphene, and hence they are directly proportional to the electron/hole excitation probability densities on the graphene atomic lattice. At filling $\nu = -2$, dI/dV map of electron excitations only shows graphene's honeycomb lattice, while at partial fillings between $\nu = -2$ and -1 the dI/dV maps of hole excitations show sublattice polarization. A key feature of graphene's ZLL is that the electron states at the K or K' valleys correspond to the A or B sublattice sites, respectively[2,39]. Therefore, the sublattice polarization observed in these maps, for example for hole excitation at $\nu = -1$, is indicative of valley polarization in the ZLL, which agrees with the expectation of a spin and valley polarized ground state |K'↑> at quarter-filling[40]. The electron excitation at this filling shows partial polarization of the orthogonal state comprising of |K'↓>, |K↑>, K>↓>. Our measurements at fillings $-2 < \nu < -1$ indicate that the ground state in this range also remains valley polarized, thereby demonstrating that FQH states in this filling range are single-component with valley symmetry breaking preceding the formation of FQH states[41].

Although valley polarization in the filling range $-2 < \nu \leq -1$ is dictated by interactions, we demonstrate that the sublattice asymmetry energy plays an important role in choosing which valley is occupied. We extract the sublattice polarization $Z = (I_A - I_B)/(I_A + I_B)$, where $I_A$ ($I_B$) are the intensity of dI/dV signals at the A (B) sublattice (supplementary materials), and plot them for the ZLL as a function of filling in Fig. 2e. Complimentary to fillings in the range $-2 < \nu \leq -1$, where we find full polarization of the hole excitation, we find for the range $1 \leq \nu < 2$, the electron excitation maps probing the unoccupied states to be fully polarized in the A sublattice. Significantly, we always find the occupied states, probed by the hole excitations, to be polarized in the B sublattice regardless of the filling factor, evident from the blue line in Fig. 2d, which is almost entirely below zero. This behavior indicates that while interactions drive the symmetry breaking, the B sublattice is favored by an apparent AB sublattice asymmetry, likely originating from partial alignment with the hBN substrate.

We turn our attention to spectroscopic imaging at charge neutrality to show that electron interactions induce an intervalley coherent electronic state in half-filled ZLL at high fields. Spectroscopic maps of $\nu = 0$ at 6T (Fig. 3a& b, device B) show a spatially varying electronic density with a periodicity that is $\sqrt{3}$ larger than that of the graphene lattice. Such reconstruction of the unit cell, also referred to as the Kekule distortion, is expected when an intervalley coherent (IVC) phase forms. This state, which is one of the four anticipated phases at charge neutrality, has a real space electronic wavefunction with probability density at both lattice sites, with one reported sighting in STM studies of multilayer graphene samples[42]. To understand the real space patterns for electron and hole excitations of this phase, we describe its valley order using a vector on a Bloch sphere: $|\psi\rangle = cos\,(\theta/2)|K\rangle + sin\,(\theta/2)e^{i\phi}|K'\rangle$, with polar angle $\theta$ and azimuthal angle $\phi$. For states with ordering vector pointing to the poles ($\theta = 0, 180°$), electron densities correspond to full valley and sublattice polarization, forming a CDW state. In contrast, when the ordering vector lies along the equator of the Bloch's sphere ($\theta = 90°$), we have equal weight on both sublattice sites, with the azimuthal angle $\phi$ characterizing the phase coherence of the wavefunctions between the two sublattices. Computing the probability density $\langle\psi|\psi\rangle$, we find that IVC state as described by $\phi = 0°$ and $180°$ (Fig. 3c) reproduces the Kekule patterns seen experimentally for electron and hole excitation in Fig. 3a&b. Naturally, the hole excitation has an orthogonal real space structure and valley polarization to the electron excitation of the same state.

More detailed analysis of the ordering vector as a function of the magnetic field reveals a continuous quantum phase transition between IVC Kekule phase and valley and sublattice polarized CDW state. We study this transition by extracting the ordering vector's polar angle $\theta$ from the Fourier transforms of real space dI/dV maps and examined it as a function of the magnetic field. With increasing field, $\theta$ shows a continuous transition from the CDW phase ($\theta=0$) to to an IVC state with $\theta$ approaching 90° in both devices (Fig. 3d). A critical field (2.2T for device C) can be identified where $\theta$ becomes non-zero while inter valley coherence emerges, as detected by the appearance of Kekule wavevectors in the FFT of dI/dV Maps. We find that both the critical field and $\theta$ at 6T measured in the two devices correlate with the influence of sublattice asymmetry imposed by the hBN substrate. The less aligned sample (device B, 13° misalignment between graphene and hBN lattice), with smaller sublattice asymmetry, shows a smaller critical field and approaches a pure IVC state with $\theta = 90°$ at a relatively lower field. This behavior is consistent with the competition between the AB sublattice asymmetry, which favors one sublattice over the other, and valley anisotropy induced by short range electron-electron and electron-phonon interactions[18], which favors valley polarization of $\theta = 90°$. The magnetic field controls the strength of the interactions and in turn the valley anisotropy energy, thereby tuning $\theta$ like the order parameter of a continuous phase transition, a behavior well captured by a mean-field description (dashed lines in Fig. 3d, supplementary materials).

Finally, we show that measurements of the spatial variation of the ordering vectors in the IVC phase can be used to directly visualize the presence of topological excitations in this state. The spatial variations are extracted by performing local Fourier analysis on the dI/dV maps, where large areas of the sample show spatially independent $\theta$ and a constant gradient for $\phi$.

Uniform gradients in $\phi$ are expected in the presence either of strain or dilute short range disorder[23]. However, near charged defects on the graphene surface, likely due to atomic adsorbates, such as that shown in Fig. 4a, we see dramatically different behavior. Near this defect, we find $\phi$ displays a swirl-like spatial variation (Fig. 4b), while the variation of $\theta$ plotted as sublattice polarization $Z = cos(\theta)$ (Fig. 4c) displays a dipole-like feature. Analysis of higher resolution electron excitation maps near this defect (Fig. 4d) shows the variations of $\phi$ (with the linear gradient background subtracted) and Z more clearly close to the defect. (Fig. 4f&g) A visual representation of the valley ordering vector texture near this defect is shown in Fig. 4e. This valley texture is consistent with that predicted for a canted anti-ferromagnetic (CAF) skyrmion excitation of the Kekule phase[22]. This topological excitation forms when the valley polarization of one spin species flips by 180° at its center, while the other spin species is devoid of any valley texture. The two key signatures of this skyrmion excitation are the dipole behavior in Z, which is equivalent to a meron-anti-meron pair (Fig. 4i), accompanied with a perpendicular oriented dipole in $\phi$ relative to the Z dipole (Fig. 4h). Simulating the valley texture using the non-linear sigma model (NLSM) (details in supplementary materials), we find excellent agreement between the results from the model calculations (Fig. 4h&i) and our experimental results (Fig. 4f&g). This CAF skyrmion carries an electric charge of $\pm e$, which is likely what caused their localization near a charged defect of the opposite sign. Our experiments show that besides the CAF skyrmion other types of valley textures are also possible (Fig. S5). It also shows that further work can map the zoo of predicted topological excitation in this and other QHFM phases of graphene[21,22]. From a broader perspective, the microscopic approach to studying valley ordering can be applied to other two-dimensional systems, such as twisted bilayer graphene.

**References:**

1.      Ezawa, Z. F. *Quantum hall effects: Field theoretical approach and related topics*. (World

Scientific, 2008).

2. Halperin, B. I. & Jain, J. K. *Fractional quantum Hall effects : new developments*. (World Scientific, 2020).

3. Nuckolls, K. P. *et al.* Strongly correlated Chern insulators in magic-angle twisted bilayer graphene. *Nature* **588**, 610–615 (2020).

4. Das, I. *et al.* Symmetry-broken Chern insulators and Rashba-like Landau-level crossings in magic-angle bilayer graphene. *Nat. Phys.* **17**, 710–714 (2021).

5. Saito, Y. *et al.* Hofstadter subband ferromagnetism and symmetry-broken Chern insulators in twisted bilayer graphene. *Nat. Phys.* **17**, 478–481 (2021).

6. Park, J. M., Cao, Y., Watanabe, K., Taniguchi, T. & Jarillo-Herrero, P. Flavour Hund's coupling, Chern gaps and charge diffusivity in moiré graphene. *Nature* **592**, 43–48 (2021).

7. Choi, Y. *et al.* Correlation-driven topological phases in magic-angle twisted bilayer graphene. *Nature* **589**, 536–541 (2021).

8. Wu, S., Zhang, Z., Watanabe, K., Taniguchi, T. & Andrei, E. Y. Chern insulators, van Hove singularities and topological flat bands in magic-angle twisted bilayer graphene. *Nat. Mater.* **20**, 488–494 (2021).

9. Feldman, B. E. *et al.* Observation of a nematic quantum Hall liquid on the surface of bismuth. *Science* **354**, 316–321 (2016).

10. Randeria, M. T. *et al.* Ferroelectric quantum Hall phase revealed by visualizing Landau level wavefunction interference. *Nature Physics* **14**, 796–800 (2018).

11. Sondhi, S. L., Karlhede, A., Kivelson, S. A. & Rezayi, E. H. Skyrmions and the crossover from the integer to fractional quantum Hall effect at small Zeeman energies. *Phys. Rev. B* **47**, 16419 (1993).

12. Nagaosa, N. & Tokura, Y. Topological properties and dynamics of magnetic skyrmions. *Nature Nanotechnology* **8**, 899–911 (2013).

13. Zhou, H., Polshyn, H., Taniguchi, T., Watanabe, K. & Young, A. F. Solids of quantum Hall skyrmions in graphene. *Nature Physics* **16**, 154–158 (2020).

14. Chatterjee, S., Bultinck, N. & Zaletel, M. P. Symmetry breaking and skyrmionic transport

in twisted bilayer graphene. *Phys. Rev. B* **101**, 165141 (2020).

15. Chatterjee, S., Ippoliti, M. & Zaletel, M. P. Skyrmion Superconductivity: DMRG evidence for a topological route to superconductivity. *arXiv* 2010.01144 (2020).

16. Khalaf, E., Chatterjee, S., Bultinck, N., Zaletel, M. P. & Vishwanath, A. Charged skyrmions and topological origin of superconductivity in magic-angle graphene. *Sci. Adv.* **7**, (2021).

17. Checkelsky, J. G., Li, L. & Ong, N. P. Zero-Energy State in Graphene in a High Magnetic Field. *Phys. Rev. Lett.* **100**, 206801 (2008).

18. Kharitonov, M. Phase diagram for the ν=0 quantum Hall state in monolayer graphene. *Phys. Rev. B* **85**, 155439 (2012).

19. Young, A. F. *et al.* Tunable symmetry breaking and helical edge transport in a graphene quantum spin Hall state. *Nature* **505**, 528–532 (2014).

20. Zhou, H. *et al.* Strong-Magnetic-Field Magnon Transport in Monolayer Graphene. *arXiv:* 2102.01061 (2021).

21. Lian, Y. & Goerbig, M. O. Spin-valley skyrmions in graphene at filling factor ν=-1. *Phys. Rev. B* **95**, 245428 (2017).

22. Atteia, J., Lian, Y. & Goerbig, M. O. Skyrmion zoo in graphene at charge neutrality in a strong magnetic field. *Phys. Rev. B* **103**, 035403 (2021).

23. Hou, C. Y., Chamon, C. & Mudry, C. Deconfined fractional electric charges in graphene at high magnetic fields. *Phys. Rev. B* **81**, 075427 (2010).

24. Miller, D. L. *et al.* Observing the quantization of zero mass carriers in graphene. *Science* **324**, 924–927 (2009).

25. Li, G., Luican, A. & Andrei, E. Y. Scanning tunneling spectroscopy of graphene on graphite. *Phys. Rev. Lett.* **102**, 176804 (2009).

26. Walkup, D. *et al.* Tuning single-electron charging and interactions between compressible Landau level islands in graphene. *Phys. Rev. B* **101**, 035428 (2020).

27. Song, Y. J. *et al.* High-resolution tunnelling spectroscopy of a graphene quartet. *Nature* **467**, 185–189 (2010).

28.	Miller, D. L. *et al.* Real-space mapping of magnetically quantized graphene states. *Nat. Phys.* **6**, 811–817 (2010).

29.	Jung, S. *et al.* Evolution of microscopic localization in graphene in a magnetic field from scattering resonances to quantum dots. *Nat. Phys.* **7**, 245–251 (2011).

30.	Luican, A., Li, G. & Andrei, E. Y. Quantized Landau level spectrum and its density dependence in graphene. *Phys. Rev. B* **83**, 041405 (2011).

31.	Chae, J. *et al.* Renormalization of the graphene dispersion velocity determined from scanning tunneling spectroscopy. *Phys. Rev. Lett.* **109**, 116802 (2012).

32.	Luican-Mayer, A. *et al.* Screening Charged Impurities and Lifting the Orbital Degeneracy in Graphene by Populating Landau Levels. *Phys. Rev. Lett.* **112**, 036804 (2014).

33.	Ghahari, F. *et al.* An on/off Berry phase switch in circular graphene resonators. *Science* **356**, 845–849 (2017).

34.	Gutiérrez, C. *et al.* Interaction-driven quantum Hall wedding cake–like structures in graphene quantum dots. *Science* **361**, 789–794 (2018).

35.	Papić, Z., Mong, R. S. K., Yazdani, A. & Zaletel, M. P. Imaging Anyons with Scanning Tunneling Microscopy. *Phys. Rev. X* **8**, 011037 (2018).

36.	Eisenstein, J. P., Pfeiffer, L. N. & West, K. W. Coulomb barrier to tunneling between parallel two-dimensional electron systems. *Phys. Rev. Lett.* **69**, 3804–3807 (1992).

37.	Dial, O. E., Ashoori, R. C., Pfeiffer, L. N. & West, K. W. High-resolution spectroscopy of two-dimensional electron systems. *Nature* **448**, 176–179 (2007).

38.	Dial, O. E., Ashoori, R. C., Pfeiffer, L. N. & West, K. W. Anomalous structure in the single particle spectrum of the fractional quantum Hall effect. *Nature* **464**, 566–570 (2010).

39.	Neto, A. H. C., Guinea, F., Peres, N. M. R., Novoselov, K. S. & Geim, A. K. The electronic properties of graphene. *Rev. Mod. Phys.* **81**, 109 (2009).

40.	Young, A. F. *et al.* Spin and valley quantum Hall ferromagnetism in graphene. *Nat. Phys.* **8**, 550–556 (2012).

41.	Dean, C. R. *et al.* Multicomponent fractional quantum Hall effect in Â graphene. *Nat.*

*Phys.* **7**, 693–696 (2011).

42. Li, S.-Y., Zhang, Y., Yin, L.-J. & He, L. Scanning tunneling microscope study of quantum Hall isospin ferromagnetic states in the zero Landau level in a graphene monolayer. *Phys. Rev. B* **100**, 085437 (2019).

**Acknowledgments:** We thank B. I. Halperin, A. H. Macdonald, N. P. Ong, and P. Kim for helpful discussions. **Funding:** The experimental work was primarily supported by NSF-DMR-1904442 and ONR-N00014-21-1-2592. Other support for the experimental efforts was provided by the Gordon and Betty Moore Foundation's EPiQS initiative grants GBMF9469, NSF-MRSEC through the Princeton Center for Complex Materials NSF-DMR-2011750, and DOE-BES grant DE-FG02-07ER46419, and the Princeton Catalysis Initiative. K.W. and T.T. acknowledge support from the Elemental Strategy Initiative conducted by the MEXT, Japan, grant JPMXP0112101001, JSPS KAKENHI grant JP20H00354, and the CREST (JPMJCR15F3), JST. M.Z. was supported through the Army Research Office through the MURI program (grant number W911NF-17-1-0323). A.Y. acknowledge the hospitality of the Aspen Center for Physics, which is supported by National Science Foundation grant PHY-1607611, and Trinity College, Cambridge UK, where his stay was supported by a QuantEmX grant from ICAM and the Gordon and Betty Moore Foundation through Grant GBMF9616. K.W. **Author contributions:** X.L., G.F., C.C. and A.Y. designed the experiment. G.F., X.L. and C.C. fabricated the sample. X.L. G.F. and C.C. performed the measurements and analyzed the data. M.Z., Z.P. and X.L. conducted the theoretical analysis. X.L., G.F., C.C., A.Y. and M.Z. wrote the mauscript with input from all authors.

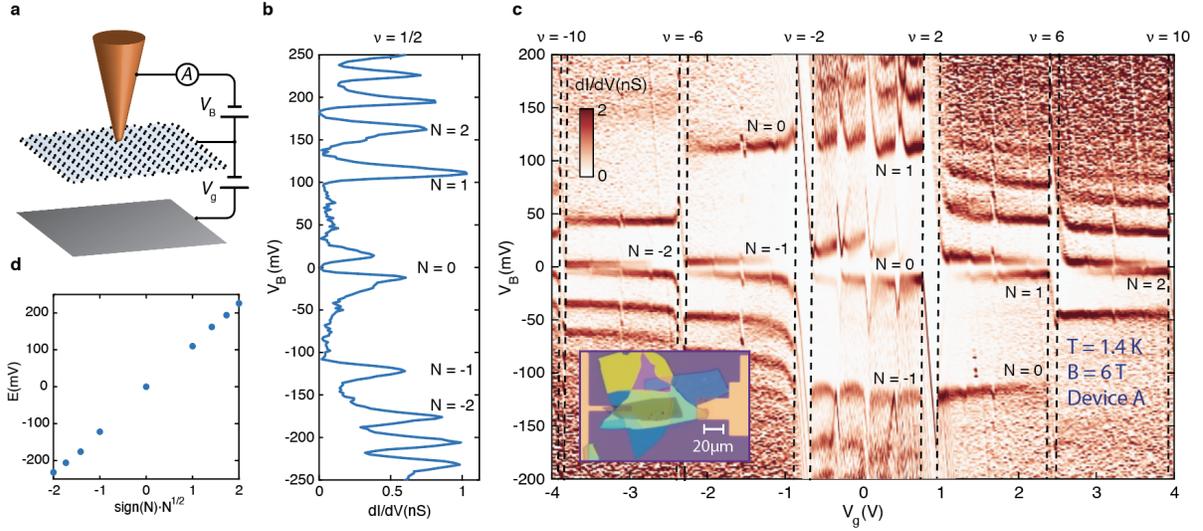

**Fig. 1. Experiment setup and large gate range spectra. a,** Schematic of the STM measurement setup. **b,** Spectrum of device A at $\nu = 1/2$ showing LL peaks of different orbital number N. **c,** Tunneling spectra of device A as a function of bias voltage and gate voltage measured at B = 6T, T = 1.4K at a fixed tip height. Inset, Optical image of device A. The left gold pad contacts the graphite gate, right contact connects with graphene. **d,** The energy of LLs taken from panel b, which matches well with $E_N = \hbar \omega_c \sqrt{N}$.

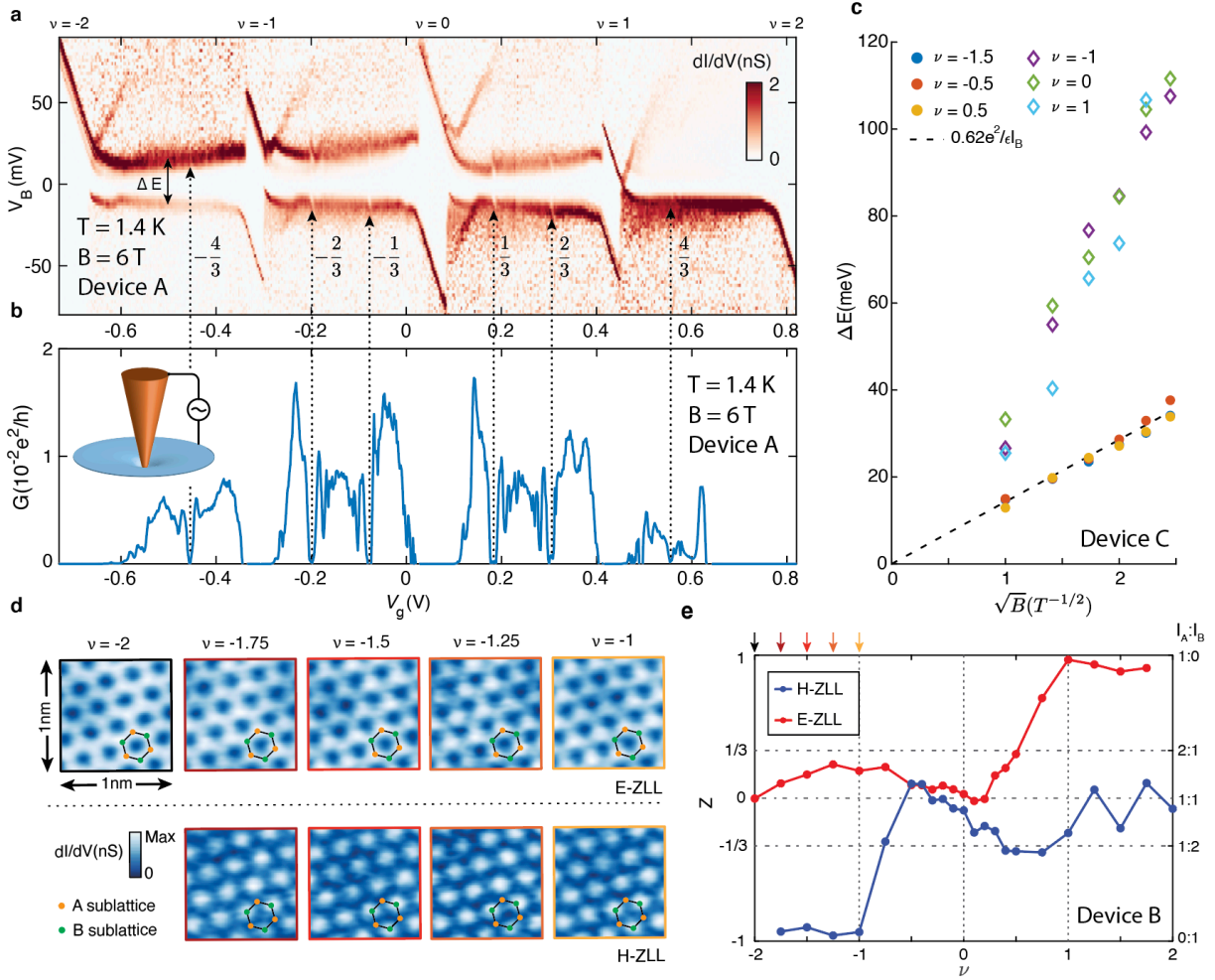

**Fig. 2. Symmetry breaking and fractional quantum Hall states of the zeroth Landau level.**
**a,** Tunneling spectrum of the zeroth Landau level between $\nu$ = -2 and 2 measured in device A. **b,**
Corbino transport measurement done on device A when contacting graphene with the tip by
reducing tip height by 2nm (B = 6T, T = 1.4K). Fractional states are detected to the 4/9$^{th}$ state.
The gate voltage at which fractional features appear coincides with the tunneling measurement in
panel a. **c,** The separations of the split ZLL peak as a function of $\sqrt{B}$, measured on device C. The
splitting at half fillings scale with Coulomb energy (black dashed line). **d,** dI/dV maps taken on
the electron excitation of the ZLL (E-ZLL) and the hole excitation of the ZLL (H-ZLL) peaks at
quarter fillings between $\nu$=-2 and -1 in device B. The hexagon pattern is the underlying graphene atomic lattice. The H-ZLL peak is fully sublattice polarized in this filling range. **e,** Sublattice polarization Z as a function of filling factors for H-ZLL and E-ZLL peak extracted by Fourier transformation of dI/dV maps.

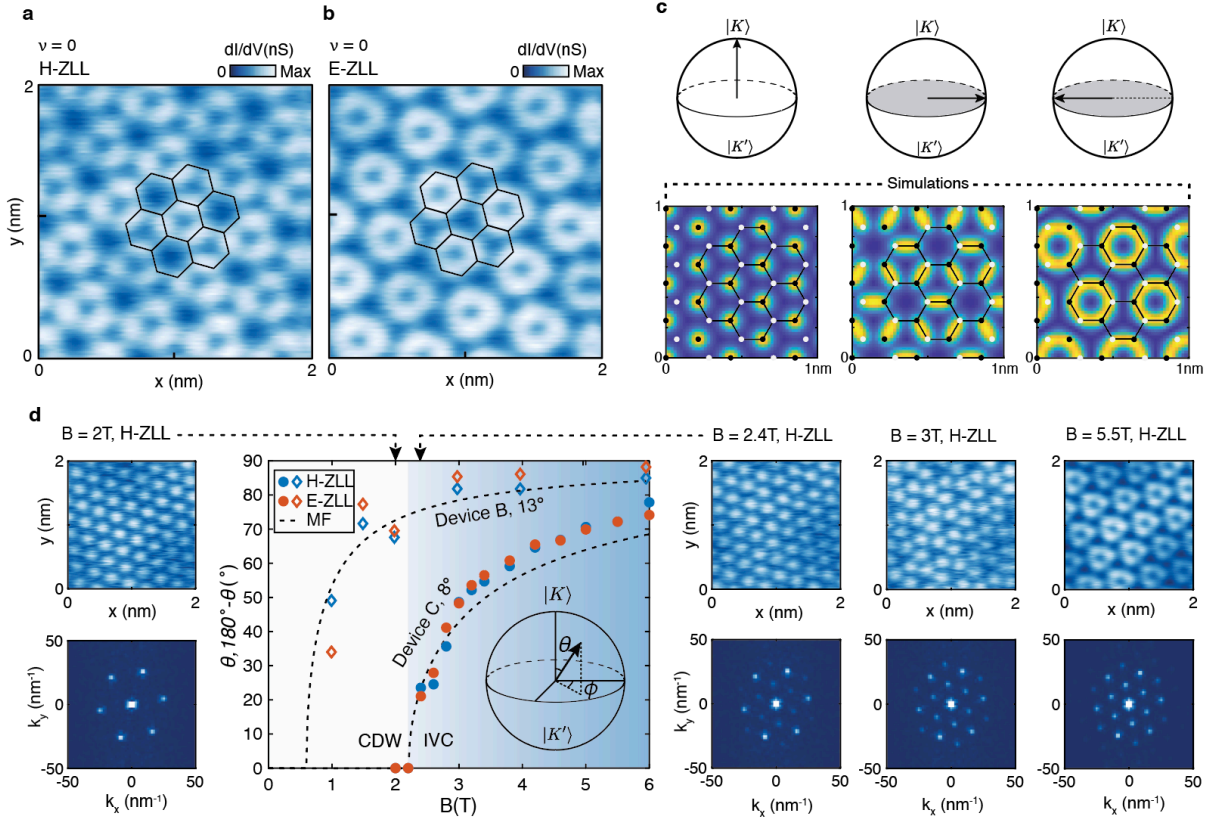

**Fig. 3 Inter-valley coherent state at CNP. a, b,** dI/dV maps at charge neutral point measured at B = 6T in device B. The hexagons represent the graphene lattice. The dI/dV maps show a Kekule reconstruction that triples the area of the unit cell. **c,** Bloch sphere plot and corresponding simulated probability density of valley polarization for CDW (left), IVC with $\phi$ of 0° (middle) and 180° (right). **d,** Polar angle $\theta$ as a function of the magnetic field in devices B and C extracted from dI/dV maps. For the E-ZLL (H-ZLL) peaks, $\theta$ (180°- $\theta$)is being plotted. The complementary behavior of H-ZLL and E-ZLL peaks confirms their orthogonal nature. The mean field (MF) behavior for $\theta$ is shown as dashed line, with critical fields of 2.2T (device C) and 0.6T (device B). The top panels on the sides show the dI/dV maps of the H-ZLL at a few representative magnetic fields. The corresponding bottom panels show the Fourier transform of the dI/dV maps above it. At B=2T, only Fourier peaks of the graphene lattice are visible, while at B= 2.4T Fourier peaks of the Kekule pattern appear and increase in intensity with increasing magnetic field.

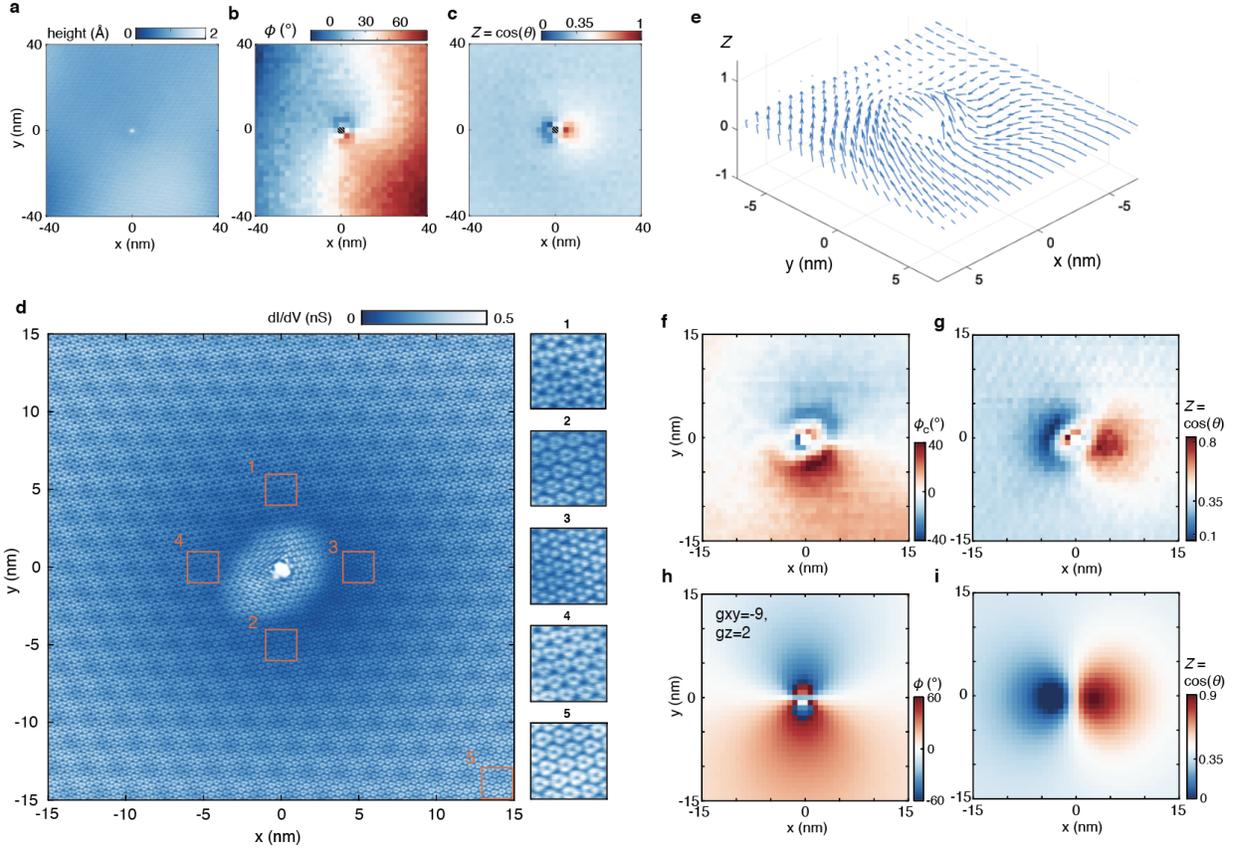

**Fig. 4 Valley skyrmion of IVC state near a charged defect. a,** Topography of the point defect found on device C. **b, c,** Azimuthal angle $\phi$ and Z polarization extracted from dI/dV maps of E-ZLL peak (Supplementary materials). **d,** dI/dV map of the E-ZLL zoomed in the area near the defect shown in panel a. Side panel shows magnified images of a few representative areas with matching labels. **e,** Valley texture extracted from panel d, visualized by arrays of arrows representing valley polarization in a Bloch sphere at each point. **f, g,** Azimuthal $\phi$ and Z polarization extracted from panel d. A linear background is subtracted from $\phi$ to produce $\phi_c$. **h, i,** Azimuthal angle and Z polarization extracted from a simulated map of electron density computed for a canted-antiferromagnetic (CAF) skyrmion using the same Fourier procedure (see supplementary information for details of the calculations).

# Supplementary Materials for

# Visualizing Broken Symmetry and Topological Defects in a Quantum Hall Ferromagnet

## 1   Sample preparation

Samples in this work were fabricated using a mechanical transfer technique. Our pickup stamp is made from a polyvinyl alcohol (PVA) coated transparent tape, which covers a polydimethyl-siloxane (PDMS) block on a glass slide handle. We used two different methods for picking up the isolated flakes. Sample A was made similarly to the previously reported TDBG device (1). Device B and C were fabricated with the pick-up order reversed: we first pick up monolayer graphene directly with PVA, and use graphene to pick up the bottom layers, hBN and graphite respectively to serve as the backgate. The stack was then transferred onto a SiO2/Si substrate with pre-patterned gold contact after dissolving the PVA film with water. After fabrication, all three samples were rinsed in water, n-methyl-2-pyrrolidone (NMP), and acetone in order to dissolve polymer residue from the surface. The devices were eventually baked in UHV at 400C overnight before transferring to the STM chamber.

## 2   STM measurements

The experiment is done in a home built UHV STM operating at T = 1.4 K. All data shown unless specified otherwise are taken at B = 6T. The measurements are performed with a tungsten tip prepared and characterized on a Cu(111) single crystal. Through controlled indentation, we shape the tip until its poke mark is confined and its spectrum features the Cu(111) surface state at the right energy. We then locate the graphene sample with a capacitance guiding technique (2).

In our spectroscopy measurements, the tip is grounded and the tunneling current is measured from the tip. Bias voltage $V_B$ is applied to the graphene sample while $V_B + V_g$ is applied to the gate to achieve a gate voltage of $V_g$ relative to the graphene sample. Differential tunneling conductance dI/dV in Fig. 1c and Fig. 2a is obtained by taking a numerical derivative of tunneling current I with respect to $V_B$ while keeping $V_g$ constant. Other differential tunneling conductances shown in this study are measured using the lock-in method with 1∼2mV AC modulation at 4 kHz. Normally, scanning tunneling spectroscopy is done at a certain setpoint voltage and setpoint current. However, as we tune the sample to different densities with the gate, measuring with the same current setpoint changes the tip height, leading to setpoint effects. To circumvent this issue, all the gate-tuned spectroscopy shown in this study are measured with a constant tip height. In principle, this can be done with the feedback turned off for the entire gate-tuned spectra. However these datasets typically take a few hours to acquire, during which the tip would drift by a significant amount. To overcome this, we first record the current at a setpoint voltage with the feedback turned off. Then we measure the spectra with feedback off but use the prerecorded constant height current to adjust tip drift compensation. This way we obtain the entire gate-tuned spectra with a constant tip height. The tip height for these measurements is set at $\nu$=1/2, $V_{set} = -0.4V$, $I_{set} = 2nA$.

We also employ constant height in small scale (<5nm) spectroscopic imaging. Since each image is obtained in only a couple of minutes, tip drifting is not a problem for this measurement. We first adjust the tilt of the scan plane locally over the scan window, using the tilt correction feature in Nanonis software. And dI/dV images are obtained by turning off the feedback and scanning the preset plane at the bias voltage corresponding to the peak we are interested in. In bigger scale spectroscopic imaging, such as the one shown in Fig. 4d, we use multipass scan function. On the first pass of each line, a topography profile was obtained with constant current at low set point ( $V_{set} = -0.4V$, $I_{set} = 100pA$ ) to avoid setpoint effects. Then on the second pass, we lower the tip by $\sim$200pm from the profile trace of the first pass and adjust $V_B$ to desired value to obtain the dI/dV.

# 3  Tip contact Corbino transport measurements

For the transport measurement, we lower the tip into the sample from the tunneling condition by a few nanometers. We found this does not damage the graphene but only changes the tip shape. The measurement circuit is similar to tunneling measurement but with a lower-gain current preamplifier. The differential tip contact Corbino conductance G is measured by lockin at 137$\sim$144Hz, 0.1$\sim$2mV oscillation.

# 4  Correlation between spectral gap and bulk transport properties in incompressible states

In Fig. S2, we compare a spectral measurement with a Corbino transport measurement as a function of gate voltage and bias voltage. We find for incompressible states at $\nu = 0, \pm 1, \pm 2$, both tunneling measurement and transport measurement shows the same large gap in bias voltage. Inside these gaps, bulk transport resistance is larger than 1GOhm. And when the bias voltage exceed the threshold, the bulk conductance is suddenly turned on. The mechanism behind such transport behavior is unclear. We emphasis that the bulk is highly conductive at fillings other than incompressible states. And the bulk is also highly conductive at the LL peaks inside incompressible states, which are above the threshold.

# 5  Numerical calculation of the LDOS

Numerical simulations of the LDOS were performed using the standard formalism for many electrons in a single Landau level on the surface of a sphere (3). The symmetries of the electron

system – the total $z$-projection of both spin and angular momentum of the electrons – were explicitly resolved. The effective interaction between the electrons is modeled according to Ref. (4), and it includes the dielectric constant (for one-sided hBN) $\epsilon_{\text{hBN}} = (1 + \sqrt{\epsilon^{\|}\epsilon^{\perp}})/2 \approx 2.5$, with $\epsilon^{\perp} = 3.0$ and $\epsilon^{\|} = 5.33$, as well as the screening by the filled Dirac sea at the RPA level (5). The screening by the gate can be neglected as the gate is placed more than $6\ell_B$ away from the sample. We fix the magnetic field to $B = 6\text{T}$, which is assumed to be strong enough to fully spin-polarize the ground state.

We evaluate the LDOS by computing the spectral functions (6, 7)

$$A_+(\epsilon) \quad = \sum_E \delta(\epsilon - (E - E_N)) |\langle N+1, E|\psi^\dagger(0)|N\rangle|^2, \tag{1}$$

$$A_-(\epsilon) \quad = \sum_E \delta(\epsilon + (E - E_N)) |\langle N-1, E|\psi(0)|N\rangle|^2, \tag{2}$$

where $E_N$ is the energy of the ground state $|N\rangle$ at particle number $N$. In principle, $A_\pm$ also depend on the spin of the removed or added electron, which in the numerical simulations can be resolved explicitly. If $\Delta_+ = E_{N+1} - E_N$ and $\Delta_- = E_{N-1} - E_N$, the thermodynamic charge gap is $\Delta = \Delta_+ + \Delta_-$. The chemical potential is defined by $\Delta_+ - \mu(N) = \Delta_- + \mu(N)$, or $\mu(N) = (\Delta_+ - \Delta_-)/2$. $A_+$ has support for $\epsilon > \Delta_+$ and $A_-$ has support for $\epsilon < -\Delta_-$.

The spectral functions in Eqs. (1)-(2) obey a number of sum rules (7, 8). For example, the zeroth-moment sum rules give the density:

$$2\pi\ell_B^2 \int A_+(\epsilon)\, d\epsilon = \langle \nu|\psi(0)\psi^\dagger(0)|\nu\rangle = 1 - \nu, \tag{3}$$

$$2\pi\ell_B^2 \int A_-(\epsilon)\, d\epsilon = \langle \nu|\psi^\dagger(0)\psi(0)|\nu\rangle = \nu, \tag{4}$$

where $|\nu\rangle$ denotes the ground state at filling $\nu$.

Explicit evaluation of Eqs. (1)-(2) is impractical due to a sum over (in principle, all) eigenstates $|E, N \pm 1\rangle$ in the spectrum. Consequently, we used a Kernel Polynomial Method (KPM) (9) which allows to iteratively evaluate LDOS by applying a Chebyshev expansion. Using KPM

expansion into a Chebyshev basis of size $\sim 100$ with the Jackson kernel from Ref. (9), in Fig. **S3** we show the resulting energy-resolved LDOS (middle panel) and its integrated version (right panel) for a system of $N_e = 14$ electrons on a sphere threaded by total magnetic flux $N_\Phi = 27$, i.e., near $\nu = 1/2$. This corresponds to filling $\nu = -3/2$ in experiment, where we assume one sublevel (e.g., K↓) is half filled, while the remaining three sublevels (K↓, K↑, K′↓) of the $N = 0$ graphene Landau level are empty. In order to account for the possibly non-uniform ground state (i.e., with angular momentum $L > 0$ on the sphere), the LDOS was explicitly averaged over different magnetic orbitals $m = 0, 1, 2, \ldots, N_\Phi$ of the added or removed electron. In Fig. **S3** we observe the LDOS displays two broad peaks located at $-0.22E_C$ and $0.36E_C$ (the peaks are identified with the median of the curves), resulting in a gap of $0.58E_C$ or approximately 32meV at 6T.

## 6 Extracting valley polarization and intervalley coherence phase from Fourier analysis of conductance maps

Focusing on one spin species (or when assuming a spin-singlet state in which the two spins are equivalent), the IQHE state is described by a spinor $(\psi_+, \psi_-) = (\cos(\theta/2), \sin(\theta/2)e^{i\phi})$ in the valley space. $\phi$ is the phase of the intervalley coherent (IVC) order while $\theta$ describes the degree of valley polarization, which we may also express as $Z = |\psi_+|^2 - |\psi_-|^2$. We would like to extract these angles from STM images of the occupied/empty density, which is in proportion to the current $I(V, \mathbf{r})$ for bias-voltages just above/below the top/bottom peak. The $K$-valley orbitals are localized at sites $\mathbf{R}_{A,i} = \mathbf{R}_i + \mathbf{r}_A$, while the $K'$ sites localized at $\mathbf{R}_{B,i} = \mathbf{R}_i + \mathbf{r}_B$. We let $w_A(\mathbf{r})$ denote the wavefunction for the $A$ orbitals, with Fourier transform $\int e^{-i\mathbf{q}\mathbf{r}}|w_A(\mathbf{r})|^2 = F(\mathbf{q})$. For $B$, $C_2$ symmetry implies $w_B(\mathbf{r}) = w_A(-\mathbf{r})$. Since sublattice and valley are locked in the $N = 0$ LL of MLG, it will be convenient to let $\tau = \pm$ denote $(A/K)$ vs $(B/K')$ together.

An electron in orbital $m$, valley $\tau$ of the $N = 0$ LL thus has an ansatz real-space wavefunc-

tion

$$\varphi_{\tau,m}(\mathbf{r}) = \sum_{\mathbf{R}_{\tau,i}} \varphi_m(\mathbf{r}) e^{i\tau \mathbf{K} \cdot \mathbf{R}_{\tau,i}} w_\tau(\mathbf{r} - \mathbf{R}_{\tau,i}) \tag{5}$$

Here $\varphi_m(z) \sim z^m e^{-\frac{1}{4\ell_B^2}|z|^2}$ are the lowest LL wavefunctions. The density of a uniform IQHE state can then be obtained using

$$n(\mathbf{r}) = \sum_{m,\tau,\tau'} \psi_\tau^* \psi_{\tau'} \varphi_{\tau,m}^*(\mathbf{r}) \varphi_{\tau',m}(\mathbf{r}) \tag{6}$$

$$= \frac{1}{2\pi\ell_B^2} \sum_{\tau,\tau',i,j} \psi_\tau^* \psi_{\tau'} e^{-i\mathbf{K} \cdot (\tau \mathbf{R}_{\tau,i} - \tau' \mathbf{R}_{\tau',j})} w_\tau^*(\mathbf{r} - \mathbf{R}_{\tau,i}) w_{\tau'}(\mathbf{r} - \mathbf{R}_{\tau',j}) \tag{7}$$

where we have used the LLL completeness relation $\sum_m |\varphi_m(\mathbf{r})|^2 = \frac{1}{2\pi\ell_B^2}$.

**Valley polarization**

We first focus on the valley-diagonal contribution to the density $\tau = \tau'$. If the orbitals are tightly localized we can restrict to $i = j$ and obtain the contribution

$$n(\mathbf{r}) \ni \sum_i \left[ |\psi_+|^2 |w_A(\mathbf{r} - \mathbf{R}_{A,i})|^2 + |\psi_-|^2 |w_B(\mathbf{r} - \mathbf{R}_{B,i})|^2 \right] \tag{8}$$

In Fourier space at reciprocal vector $\mathbf{G}$,

$$n(\mathbf{G}) \ni \left[ F(\mathbf{G}) |\psi_+|^2 e^{-i\mathbf{G} r_A} + F^*(\mathbf{G}) |\psi_-|^2 e^{-i\mathbf{G} r_B} \right] \tag{9}$$

where $F(\mathbf{q})$ is the Fourier transform (form factor) of $|w_A(\mathbf{r})|^2$.

From this expression we would like to extract the degree of valley polarization $Z = |\psi_+|^2 - |\psi_-|^2$. However, under a shift of the origin by $\mathbf{R}_0$, the result transforms as $n(\mathbf{G}) \to n(\mathbf{G}) e^{-i\mathbf{G} \cdot \mathbf{R}_0}$. Since we don't a priori know where to fix the origin in a given region of the STM image, we need a way to extract order parameters in the presence of this ambiguity.

To do so, we consider products of the form $\prod_{\mathbf{q}_i} n(\mathbf{q}_i)$, where $\sum_i \mathbf{q}_i = 0$, which is thus invariant under a shift of $\mathbf{R}_0$. Let $\mathbf{G}_i = [C_3]^i \mathbf{G}_0$, $i = 0, 1, 2$, be the three $C_3$-related reciprocal vectors. The combination $\Phi = \arg n(\mathbf{G}_0) n(\mathbf{G}_1) n(\mathbf{G}_2) = 3 \arg(F(\mathbf{G}) |\psi_+|^2 e^{2\pi i/3} +$ $F^*(\mathbf{G})|\psi_-|^2 e^{-2\pi i/3}) = 3\arg(|\psi_+|^2 e^{i\alpha} + |\psi_-|^2 e^{-i\alpha})$ is thus an invariant, where $\arg F(\mathbf{G})e^{2\pi i/3} = \alpha$. Without knowing $\alpha$, we can't convert this directly to $Z$. However, if we assume $w_A(\mathbf{r}) = w_A(-\mathbf{r})$ so that $F(\mathbf{G})$ is real, $\alpha = \frac{2\pi}{3}$ or $\alpha = \frac{2\pi}{3} + \pi$, and assuming the former we find $Z = -\tan(\frac{\Phi}{3})/\sqrt{3}$. By applying this method at $\nu = -1$, where we know the sublattice splitting ensures the valley-polarized state $Z = \pm 1$, we can confirm that $\alpha \approx 0$.

**IVC (Kekule) order**

Our discussion thus far neglected possible coherence between $K$ and $K'$ coming from $\psi_+^* \psi_-$. The coherences will be peaked on the set of nearest-neighbor bonds which we denote by $< i,j >$. The valley coherent contribution at wavevector $\mathbf{K} - \mathbf{K}'$ is

$$n(\mathbf{r}) \ni \frac{1}{2\pi\ell_B^2} \sum_{<i,j>} \psi_+^* \psi_- e^{-i\mathbf{K}\cdot(\mathbf{R}_{A,i}+\mathbf{R}_{B,j})} w_A^*(\mathbf{r} - \mathbf{R}_{A,i}) w_B(\mathbf{r} - \mathbf{R}_{B,j}) + h.c. \quad (10)$$

There are three such bonds per unit cell, with centers we will denote by $\mathbf{R}_{\alpha,i}$ where $\alpha = 0, 1, 2$ denotes the types. In terms of $\mathbf{R}_{\alpha,i}$ the above expression reduces to

$$n(\mathbf{r}) = \frac{1}{2\pi\ell_B^2} \sum_{\alpha,i} \psi_+^* \psi_- e^{-i2\mathbf{K}\cdot\mathbf{R}_{\alpha,i}} B_\alpha(\mathbf{r} - \mathbf{R}_{\alpha,i}) + h.c. \quad (11)$$

where $B_\alpha(\mathbf{r}) = w_A^*(\mathbf{r} - \mathbf{R}_{A,i}) w_B(\mathbf{r} - \mathbf{R}_{B,j})$ with $B_\alpha(\mathbf{r}) = B_\alpha(-\mathbf{r})$. The Fourier transform is

$$n(\mathbf{K} + \mathbf{G}) = \psi_+^* \psi_- \sum_\alpha B_\alpha(\mathbf{K} + \mathbf{G}) e^{-i(3\mathbf{K}+\mathbf{G})\cdot\mathbf{R}_\alpha} \quad = \psi_+^* \psi_- e^{2\pi i/3} \sum_\alpha B_\alpha(\mathbf{K} + \mathbf{G}) e^{-i(\mathbf{K}+\mathbf{G})\cdot\mathbf{R}_\alpha}$$
$$(12)$$

where $B_\alpha(\mathbf{q})$ is the Fourier transform of $B_\alpha(\mathbf{r})$. To extract this in a shift-invariant manner, we note that $n(\mathbf{K}_0)n(\mathbf{K}_1)n(\mathbf{K}_2) \propto -(\psi_+^* \psi_-)^3 = -e^{3i\phi}$ where $\mathbf{K}_i$ are the $C_3$-related $K$-points and we have exploited $C_3$ to relate the $B_\alpha$. Note that by $C_2$-symmetry $B_\alpha(\mathbf{q})$ is real, and we assumed $B_\alpha(\mathbf{K}_i) > 0$, which is certainly reasonable for smooth orbitals. Regardless, if $B_\alpha(\mathbf{K}_i) < 0$, we instead obtain $(\psi_+^* \psi_-)^3$, which differs only by a redefinition $\phi \to \phi + 2\pi/6$.

# 7   Non-linear sigma model modelling of $\nu = 0$ phases

In this section we describe the non-linear sigma model (NLSM) used to predict the mean field phase diagram of the CDW - IVC transition and the spatial structure of the impurity-pinned skyrmion textures.

An IQHE ferromagnet is characterized by a $4 \times 4$ projection matrix $P^2 = P$ describing which of the 4 isospin states are filled. At neutrality, $\text{Tr}(P) = 2$, and it will be convenient to write $P = \frac{1}{2}(\mathbb{1} + Q)$ where $\text{Tr}(Q) = 0$ and $Q^2 = \mathbb{1}$. The $Q$ captures the different symmetry-broken phases. Letting $\tau, \sigma$ denote Pauli matrix in the valley/spin space, for example: (1) spin ferromagnet $Q_{\text{FM}} = \hat{n} \cdot \sigma$; (2) the valley and sublattice polarized state $Q_{\text{CDW}} = \tau^z$; (3) the "partially sublattice polarized" (alias IVC / Kekule) state $Q_{\text{PSP}} = \sin(\theta)(\cos(\phi)\tau^x + \sin(\phi)\tau^y) + \cos(\theta)\tau^z$ (4) the canted-antiferromagnet state $Q_{\text{CAF}} = \tau^z \sin(\theta)(\cos(\phi)\sigma^x + \sin(\phi)\sigma^y) + \cos(\theta)\sigma^z$. (10)

In Ref. (10) it was argued that energy per flux of a uniform IQHE state takes the general phenomenological form

$$H_{\text{MF}} = \frac{1}{2} \sum_{\mu=x,y,z} u_\mu \left[ \text{Tr}(\tau^\mu P)^2 - \text{Tr}(\tau^\mu P \tau^\mu P) \right] - \Delta_{AB}\text{Tr}(P\tau^z) - \Delta_Z\text{Tr}(P\sigma^z) \tag{13}$$

Here $\Delta_{AB}$ and $\Delta_Z$ are the single particle sublattice and Zeeman splittings (note the band gap is thus $2\Delta_{AB}$ in this convention). The coupling constants $u_\mu$ are not precisely known (they arise from a combination of short-range Coulomb and phonons), but should roughly scale with $B$ as $u_\mu = \frac{a}{\ell_B}E_C g_\mu$, where $a$ is the lattice spacing, $E_C = \frac{e^2}{4\pi\epsilon\ell_B}$ is the Coulomb scale, and $g_x = g_y = g_{xy}, g_z$ are some unknown $B$-independent constants. Assuming the $g_\mu$ are constant for a given dielectric environment, they can be determined by measuring the location of phase transitions in the $\Delta_Z, \Delta_{AB}, B$ plane, as we will derive below.

Based on prior experimental work (10–12) it is believed that $g_{xy} < 0$ and $g_z > 0$. As we will review, this results in the CDW phase at low-$B$ and the PSP (IVC) phase at intermediate-$B$, separated by a continuous transition at some $B_c^{\text{CDW}-\text{PSP}}$, exactly as observed in our experiment. If $g_z > -g_{xy}$, then for even larger-$B$ we predict a first order transition from the PSP to the CAF state at some $B_c^{\text{PSP}-\text{CAF}}$. This transition is not observed in our experiments up to $B = 6\,\text{T}$. Following the discussion of Ref. (10), the critical $B_c$ for these two transitions are derived below.

## Critical $B_c^{\text{CDW}-\text{PSP}}$ of the CDW - PSP transition

Plugging in the ansatz $Q = \hat{n} \cdot \tau$, the energy of the PSP state is

$$E_{\text{PSP}} = u_{xy}(n_x^2 + n_y^2) + u_z n_z^2 - 2\Delta_{AB} n_z = u_{xy} + (u_z - u_{xy})n_z^2 - 2\Delta_{AB} n_z \qquad (14)$$

The energy is minimized when the state cants to

$$n_z = \frac{\Delta_{AB}}{u_z - u_{xy}} = \frac{\Delta_{AB}}{E_C} \frac{\ell_B}{a} \frac{1}{g_z - g_{xy}} = \frac{4\pi\epsilon}{e^2 a} \frac{1}{g_z - g_{xy}} \frac{\Delta_{AB}}{eB} \qquad (15)$$

The decrease of $n_z$ as $\frac{1}{B}$ is remarkably close to our STM results (main text, Fig. 3d). Since the CDW is just a limiting case $n^z = 1$ of the PSP, the critical point for the CDW-PSP transition is thus at $n_z = 1$, or $g_z - g_{xy} = \frac{4\pi\epsilon}{e^2 a} \frac{\Delta_{AB}}{eB}$. Knowing $\Delta_{AB}$ and $B_c$ for a sample, we can then extract $g_z - g_{xy}$.

Using STM to measure the splitting of the ZLL in the $B \to 0$ limit and in $\nu = 2$ under $B = 6T$, we obtain $\Delta_{AB} = 5 \sim 10\text{meV}$ in Device C ($\theta_{BN} = 8°$), and from our analysis of the onset of the Kekule distortion, $B_c = 2.2T$. Using $\epsilon = 4\epsilon_0$ and assuming $\Delta_{AB} = 7\text{meV}$, we thus conclude $g_z - g_{xy} \sim 11$. Device B is expected to have a smaller $\Delta_{AB}$, due to a larger BN alignment angle $\theta_{BN} = 13°$, in agreement with its lower $B_c = 0.6\,\text{T}$. Unfortunately we were unable to quantitatively measure $\Delta_{AB}$ in Device B before it was compromised, so we cannot use it to independently estimate $g_z - g_{xy}$.

Within the PSP phase, the optimized energy is

$$E_{\text{PSP}} = u_{xy} - \frac{\Delta_{AB}^2}{u_z - u_{xy}} \qquad (16)$$

## Critical $B_c^{\mathrm{PSP-CAF}}$ of the PSP-CAF transition

The above measurement detects $g_z - g_{xy}$, but leaves $g_z + g_{xy}$ unconstrained. In principle it can be determined by measuring the $B_c$ of the PSP-CAF transition, should it exist. The energy of the CAF is

$$E_{\mathrm{CAF}} = -u_z + \frac{\Delta_Z^2}{2u_{xy}} \qquad (17)$$

assuming $u_{xy} < 0$. Comparing the energies of the PSP and CAF phases, the CAF is obtained at high-$B$ only if $g_z > -g_{xy}$. Based on earlier experiments which observed a transition consistent with the existence of the CAF at high fields (11, 12), this was believed to be the case, at least in samples with *two*-sided hBN encapsulation. The critical point for the PSP - CAF transition is $u_{xy} + u_z = \frac{\Delta_Z^2}{2u_{xy}} + \frac{\Delta_{AB}^2}{u_z - u_{xy}}$. At low fields, where $\Delta_Z \approx 0$, this reduces to $(u_{xy} + u_z)(u_z - u_{xy}) = u_z^2 - u_{xy}^2 = \Delta_{AB}^2$. For the typical values of $g \sim 5 - 10$ estimated from earlier experiments, obtaining a transition below $B = 6$T requires samples with smaller $\Delta_{AB}$, which may explain the absence of this transition in our experiments. Alternatively, it could be that impurities, which will couple to the IVC order $\phi$ as a random-XY field, help pin the PSP state and further favor PSP over CAF.

On the other hand, comparison of the observed defect textures with the NLSM suggests that $g_{xy} \sim -9, g_z \sim 2$, so it may be that in devices with only an hBN substrate (which are suitable for STM), rather than full encapsulation (as in the previous experiments), are in a different regime than these earlier experiments.

## NLSM solution for skyrmion textures

In the presence of defects or excitations, the projector $P(\mathbf{r})$ can vary in space, costing an additional elastic energy $H_\rho = \frac{\rho_s}{2} \int d^2 r (\nabla P)^2$, where $\rho_s = \frac{1}{16\sqrt{2\pi}} E_C$. (13) In this section, we obtain a non-linear sigma model for such textures which can be used to numerically solve for the structure of skyrmions in the presence of the anisotropies, $g_{xy}, g_z, \Delta_{AB}$. Our analysis generalizes Ref. (14) by including the effect of $\Delta_{AB}$.

In general, the space of possible rank-2 projectors $P$ is 8-dimensional at $\nu = 0$, leading to a plethora of different possible textures. (14) Fortunately, in the regime $g_{xy} < 0$ and $g_z > 0$ of interest to experiment, only a restricted subspace is energetically relevant. (14) These are states in which the skyrmions lie in the CDW/PSP/CAF manifold. Furthermore, when $\Delta_Z \sim 0$, we can ignore the canting of the CAF, in which case it reduces to a colinear AF. We may thus assume that spin-rotation symmetry is preserved about a particular spin axis (say $\hat{z}$). Under these assumptions a sufficiently general ansatz is obtained by decomposing the projector into the contribution from each spin sector $\sigma^z = \pm$ individually, $P = P^+ + P^-$. Within each spin sector, the projector $\mathrm{Tr}(P^\pm) = 1$ is fully specified by an orientation $\mathbf{n}^\pm \cdot \tau$ in the valley space. Again, this is not the most general possible ansatz - it rules out the FM phase and a non-colinear AF for example - but it does account for the energetically relevant skyrmions. For example, the spin-singlet CDW/PSP/Kekule manifold is $\mathbf{n}^+ = \mathbf{n}^-$, while the AF is $\mathbf{n}^+ = -\mathbf{n}^- = (0, 0, 1)$.

In terms of the NLSM parameters $\mathbf{n}^\pm$, the spatially uniform mean field energy is

$$E_{\mathrm{MF}} = \sum_\mu u_\mu n_\mu^+ n_\mu^- - \Delta_{AB}(n_z^+ + n_z^-) \qquad (18)$$

Thus, generalizing the NLSM model of the spinful integer quantum Hall effect, we obtain a NLSM of the form

$$H = \int \frac{d^2 r}{2\pi \ell_B^2} \left[ \frac{\rho_s}{2}((\nabla \mathbf{n}^+)^2 + (\nabla \mathbf{n}^-)^2) + \sum_{i=x/y/z} u_i n_i^+ n_i^- - \Delta_{AB}(n_z^+ + n_z^-) \right] + H_{LR}[\rho] + H_{imp}[\rho] \qquad (19)$$

where $\rho_s = \frac{E_C}{16\sqrt{2\pi}}$ is the stiffness of an IQH ferromagnet. $H_{LR}[\rho]$ is the energy coming from the long-range part of the Coulomb interaction, which depends on the charge density $\rho$ via the formula for skyrmion charge density given below. $H_{imp}[\rho]$ is the pinning impurity potential, which we take to be a Coulomb impurity.

We obtain the lowest energy $\mathbf{n}^{\pm}$ configuration by discretizing the NLSM on a $200 \times 200$ toroidal grid and applying a numerical optimization method (BFGS).

**Topological charge density.** The general formula for the skyrmion charge density is $\rho = \frac{1}{2\pi i}\epsilon^{\mu\nu}\mathrm{Tr}(P\partial_\mu P\partial_\nu P)$. For our ansatz, $P = P^+ + P^-$, it decouples spin-by-spin, $\rho^{\pm} = \frac{1}{2\pi i}\epsilon^{\mu\nu}\mathrm{Tr}(P^{\pm}\partial_\mu P^{\pm}\partial_\nu P^{\pm})$. Substituting $\partial_\mu P^{\pm} = \frac{1}{2}\partial_\mu \mathbf{n}^{\pm}\cdot\tau$ and $\partial_\mu|\mathbf{n}| = 0$, we obtain

$$\rho^{\pm} = \frac{1}{2\pi i}(\frac{1}{2})^3\epsilon^{\mu\nu}(\partial_\mu n_i^{\pm})(\partial_\nu n_j^{\pm})n_k\partial_\mu\mathrm{Tr}(\tau^i\tau^j\tau^k) \tag{20}$$

$$= \frac{1}{2\pi}(\frac{1}{2})^2\epsilon^{\mu\nu}(\partial_\mu n_i^{\pm})(\partial_\nu n_j^{\pm})n_k\epsilon^{ijk} \tag{21}$$

$$\rho^{\pm} = \frac{1}{4\pi}\mathbf{n}^{\pm}\cdot(\partial_x\mathbf{n}^{\pm})\times(\partial_y\mathbf{n}_j^{\pm}) \tag{22}$$

This gives the well-known expression equating the electrical charge with the topological charge of the skyrmion. (13) The total charge $\rho(\mathbf{r}) = \rho^+(\mathbf{r}) + \rho^-(\mathbf{r})$ then interacts via the gate-screened Coulomb interaction $H_{LR} = \frac{e^2}{4\pi\epsilon}(\frac{1}{r} - \frac{1}{\sqrt{r^2+4d^2}})$ and a gate-screened Coulomb impurity $H_{imp}$ a distance of $\ell_B$ below the sample.

**Analytic estimate of skyrmion sizes.** How large are the expected skyrmions? We can answer this question analytically if we assume that $u_i, \Delta_{AB}$ are small compared to the stiffness. In this case, we can make an approximation in which we assume that $\mathbf{n}^+(r)$ contains a skyrmion, while $\mathbf{n}^-$ remains at its constant bulk value. This neglects the "backreaction" effect in which $u_i$ causes $\mathbf{n}^-$ to adjust due to the deformation in $\mathbf{n}^+$, lowering the energy; this will thus slightly *under*-estimate the skyrmion size. Referring to the anisotropy in Eq. (19), when holding $\mathbf{n}^-$ fixed we see that the $\mathbf{n}^+$ experiences a net "Zeeman" field of $u_i n_i^- - \Delta_{AB}\hat{z}$. Plugging in the bulk PSP solution for $\mathbf{n}^-$, we find an effective Zeeman field with magnitude precisely $|u_{xy}|$. Consequently the problem reduced to that of an SO(3) spin skyrmions in a magnetic field under the replacement $\frac{1}{2}g\mu_B B \to u_{xy}$. Converting the expressions of Ref. (13) to units of $E_C$, we find that (up to constants within the log) the skyrmion size is

$$\ell_s = \ell_B \left( \frac{0.0867}{\frac{u_{xy}}{E_C} |\log(\frac{u_{xy}}{E_C})|} \right)^{1/3} \tag{23}$$

For values $g_{xy} \sim 5 - 10$meV at $B = 6$T, this gives $\ell_s = 6 - 8$nm, consistent with experiment. They are thus small skyrmions, comparable to the magnetic length. In this regime the NLSM is not expected to be quantitatively accurate, but nevertheless appears to reproduce the gross features of the experiment.

## 8 Comparison of NLSM results with experiment.

Based on the observed CDW-PSP transition, we fix $g_z - g_{xy} = 11$ and treat $g_{xy} < 0$ as a free parameter to be matched with experiment. Numerically solving for the ground state of the NLSM in the presence of a charge-$e$ skyrmion, we produce simulated experimental data as in Fig 4.

Note that our experimental procedure for extracting the IVC phase $\phi = \arg(\psi_+^* \psi_-)$ from the Fourier-transformed spectroscopy measurements (Sec.6) implicitly assumed that both spin species have the same valley order, $\mathbf{n}^+ = \mathbf{n}^-$. This is true away from defects, but not in the core of skyrmions, where $\mathbf{n}^+ \neq \mathbf{n}^-$. Because STM does not resolve by spin, the LDOS is sensitive only to the total electron density, and hence the spin-averaged valley polarization $\mathbf{n} = (\mathbf{n}^+ + \mathbf{n}^-)/2$. As a result, when we apply our procedure for extracting $\phi$ from the Fourier-transformed LDOS, we are in fact measuring the *averaged* quantity $\phi = \arg(\psi_{\uparrow,+}^* \psi_{\uparrow,-} + \psi_{\downarrow,+}^* \psi_{\downarrow,-})$.

In order to compare our NLSM result with experiment (main text, Fig. 4f-i) we thus compute $Z$ and $\phi$ for the spin-averaged NLSM field $\mathbf{n} = (\mathbf{n}^+ + \mathbf{n}^-)/2$, with $\phi = \arg(n_x + in_y)$ (Fig.4(h)) and $Z = n_z$ (Fig.4(i)). Note that for charge-$e$ skyrmions, in which only *one* of either $\mathbf{n}^+$ or $\mathbf{n}^-$ contains the texture, we do not observe a full $360°$ winding in $\phi$ precisely because of this averaging.

We choose $g_{xy}$ in order to qualitatively match theory and experiment. We find $g_{xy}$ = -6 (-11) results in a skyrmion with $\sim$ 15nm (7nm) Z polarization core separation (the distance between maximum Z polarization and minimum Z polarization). And the value $g_{xy} = -9$ produces a skyrmion with Z core separation of 10nm, similar to the experimental observation, though due to the exponent of $1/3$ which relates $g_{xy}$ and the skyrmion size (Eq.(23)) , $g_{xy}$ cannot be constrained very precisely.

# References

1. X. Liu, *et al.*, *Nature Communications* **12**, 2732 (2021).

2. G. Li, A. Luican, E. Y. Andrei, *Review of Scientific Instruments* **82**, 073701 (2011).

3. F. D. M. Haldane, *Phys. Rev. Lett.* **51**, 605 (1983).

4. F. Yang, *et al.*, *Phys. Rev. Lett.* **126**, 156802 (2021).

5. K. Shizuya, *Phys. Rev. B* **75**, 245417 (2007).

6. S. He, P. M. Platzman, B. I. Halperin, *Phys. Rev. Lett.* **71**, 777 (1993).

7. R. Haussmann, H. Mori, A. H. MacDonald, *Phys. Rev. Lett.* **76**, 979 (1996).

8. A. H. MacDonald, *Phys. Rev. Lett.* **105**, 206801 (2010).

9. A. Weiße, G. Wellein, A. Alvermann, H. Fehske, *Rev. Mod. Phys.* **78**, 275 (2006).

10. M. Kharitonov, *Physical Review B* **85**, 155439 (2012).

11. A. Zibrov, *et al.*, *Nature Physics* **14**, 930 (2018).

12. H. Zhou, *et al.*, *arXiv preprint arXiv:2102.01061* (2021).

13. S. L. Sondhi, A. Karlhede, S. A. Kivelson, E. H. Rezayi, *Phys. Rev. B* **47**, 16419 (1993).

14. J. Atteia, Y. Lian, M. O. Goerbig, *Physical Review B* **103**, 035403 (2021).

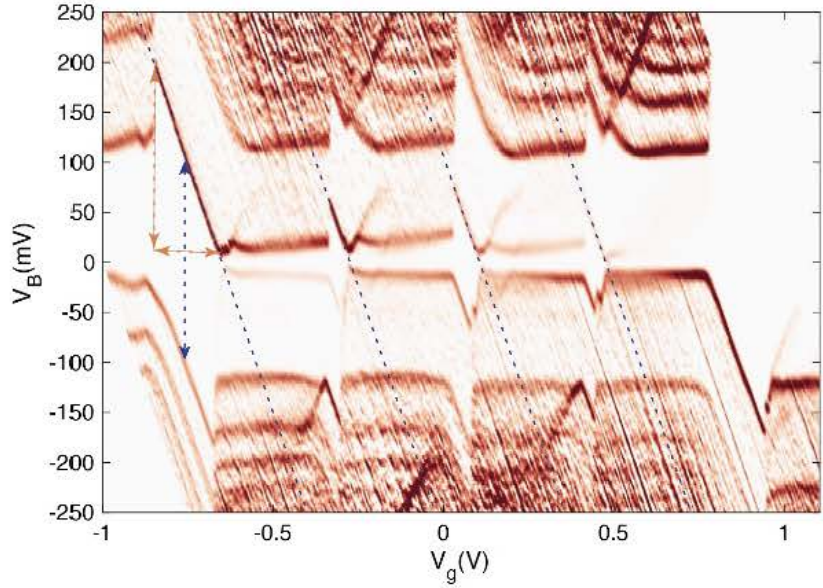

**Figure S1**: **Zoomed in Spectra between $\nu$= -2.5 and 2.5 showing the details of incompressible gaps.** The slopes of the LL peaks inside the incompressible states are $dV_B/dV_g = -1$ (the dashed black lines are guiding lines with exact -1 slope), which seem to suggest the backgate is simply moving the Fermi level in the gap and there is no in-gap impurity state. However, for single particle gaps, we found the LL peaks are discontinuous on one side of the gap, as a result of gate voltage needed to tune through the gap (horizontal orange arrowed line) is larger than the cyclotron gap itself. Equally puzzling is that the apparent gap across the Fermi energy (blue vertical arrow) is about twice as big as the cyclotron gap.

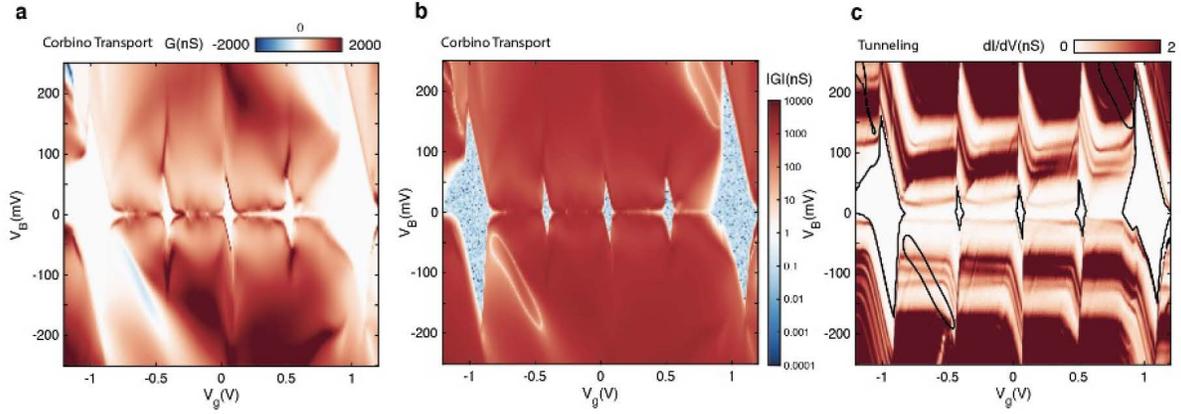

**Figure S2**: **Tip contact Corbino transport measurement at finite bias and comparison with spectral measurement. a, b,** Corbino transport measurement by plunging the tip 2nm into the sample. a(b) is plotted in the linear (logarithmic) scale. There is an insulating "diamond" at each compressible state. **c,** Spectra measured in the same area as a&b. The black lines mark the 1nS conductance contour of the transport measurement in a&b, which matches with where the tunneling conductance vanishes. This suggests the expanded tunneling gap at incompressible states is related to the bulk transport behavior of the sample. The strong peaks near ±60mV and ±80mV are phonon-assisted tunneling copies of the direct tunneling peaks at lower energies. They are more prominent in this dataset, presumably due to a blunter tip.

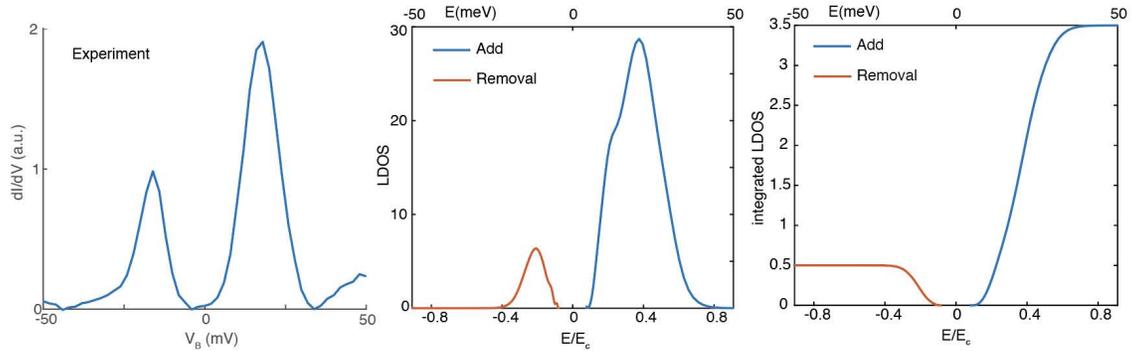

**Figure S3**: LDOS near $\nu=1/2$ computed via KPM method (9) [middle, right panels] versus the experimental $dI/dV$ spectrum at $\nu=-3/2$ [left panel]. Middle panel: LDOS resolved by the energy and averaged over all orbitals $m$ where an electron could be added or removed. The data for adding an electron includes contributions from either spin orientation, as well as the three empty sublevels of the $N=0$ graphene Landau level (see text for details). In contrast, the removed electron must have spin-up due to the ground state being fully spin-polarized. Right panel: integrated LDOS, $\int_{\mu}^{|E|} d\epsilon A_+(\epsilon)$ and $\int_{-|E|}^{\mu} d\epsilon A_-(\epsilon)$, illustrating the convergence to the zeroth sum rule in Eqs. (3)-(4).

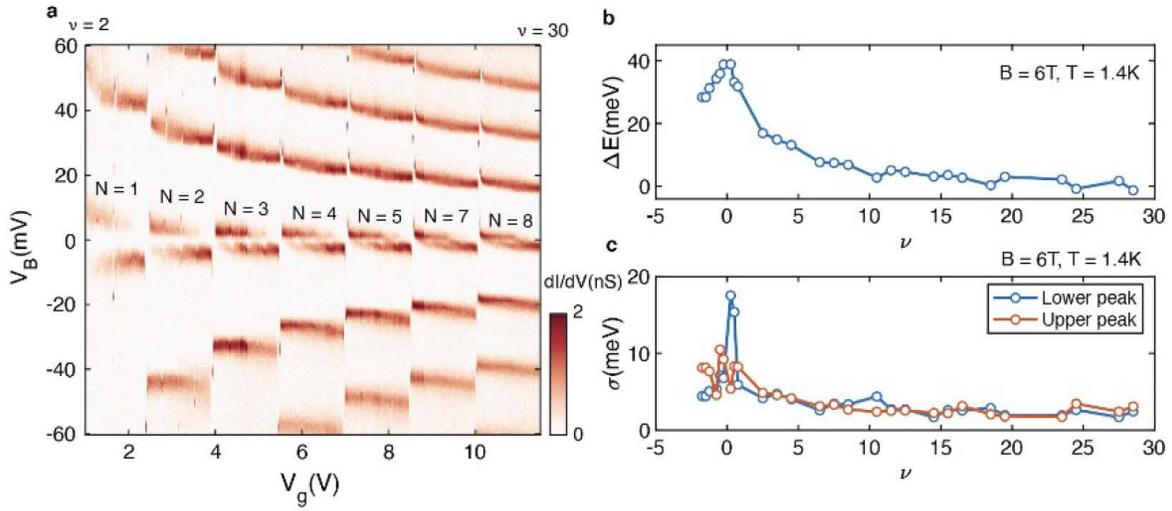

**Figure S4**: **Energy splitting across the Fermi energy at higher LLs.** **a**, Spectra at higher filling factors on the electron side. **b**, Extracted energy splitting of the LL at the Fermi level at half fillings from Fig. 2a and Fig. S2a. The split LL peaks are fitted by Gaussian functions, and $\Delta E$ is the separation between the center of the lower peak and center of the upper peak. As filling factor increases, $\Delta E$ get smaller at higher LLs, likely due to the reduction of Coulomb interaction by Landau level mixing and dielectric screening. **c**, The width of the peaks (standard deviation of the Gaussian fit) at half fillings.

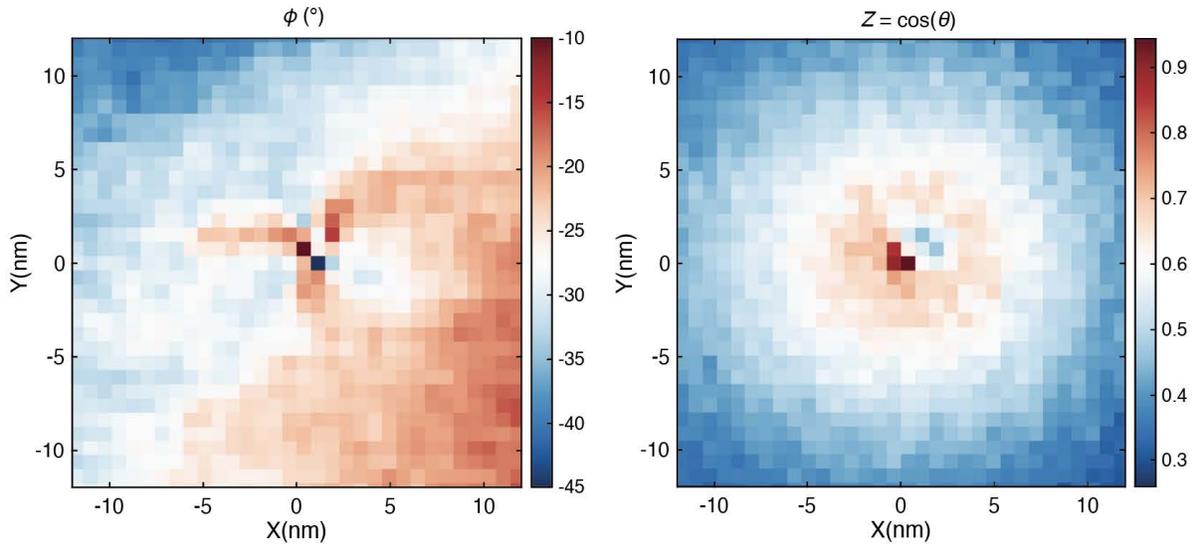

**Figure S5**: **Another observed valley textures.** Azimuthal angle $\phi$ and Z polarization of the Kekule phase at charge neutrality around a different defect extracted from E-ZLL.